\documentstyle[pre,preprint,aps]{revtex}
\tightenlines
\newcommand{\siml}{\stackrel{<}{\sim}}
\newcommand{\simg}{\stackrel{>}{\sim}}

\begin{document}
\draft

\title{
Dynamical mean-field theory of 
noisy spiking neuron ensembles: \\
Application to the Hodgkin-Huxley model
}
\author{
Hideo Hasegawa
\footnote{e-mail address:  hasegawa@u-gakugei.ac.jp}
}
\address{
Department of Physics, Tokyo Gakugei University,
Koganei, Tokyo 184-8501, Japan
}
\date{\today}
\maketitle
\begin{abstract}
A dynamical mean-field approximation (DMA) previously 
proposed by the present author 
[H. Hasegawa, Phys. Rev E {\bf 67}, 041903 (2003)]
has been extended
to ensembles described by a general 
noisy spiking neuron model.
Ensembles of $N$-unit neurons, each of which is
expressed by coupled $K$-dimensional differential equations (DEs),
are assumed to be subject to spatially correlated white noises. 
The original $KN$-dimensional 
{\it stochastic} DEs have been replaced 
by $K(K+2)$-dimensional {\it deterministic} DEs
expressed in terms of means and
the second-order moments of {\it local} and {\it global} variables:
the fourth-order contributions are taken 
into account by the Gaussian
decoupling approximation.
Our DMA has been applied to an ensemble of
Hodgkin-Huxley (HH) neurons ($K=4$), for which
effects of the noise, the coupling strength
and the ensemble size on
the response to a single-spike input have been
investigated. 
Results calculated by DMA theory
are in good agreement with those obtained by direct simulations.

\end{abstract}

\noindent
\vspace{0.5cm}
%
\section{INTRODUCTION}


It is well known that
a small cluster of cortex may contain thousands 
of similar neurons. Each neuron which receives spikes 
from hundreds of other neurons, generates spikes
propagating along the axon towards synapses
exciting neurons in the next stage.
Dynamics of an individual neuron with 
voltage-dependent ionic channels can be described by 
Hodgkin-Huxley-type (HH) model \cite{Hodgkin52},
or by reduced, simplified neuron models such as
integrate-and-fire (IF), 
FitzHugh-Nagumo (FN) \cite{FitzHugh61,Nagumo62}
and Hindmarsh-Rose (HR) models \cite{Hindmarsh82}. 
Although the response of a single neuron
in {\it vitro} is rather accurate, that in {\it vivo}
is not reliable \cite{Mainen95}.
This is due to noisy environment in living brains,
where various kinds of noises are reported to be ubiquitous
(for a review see \cite{Segund94}). In recent years, 
the population of
neuron ensembles has been recognized to
play important roles in the information transmission 
({\it pooling effect}) \cite{Hopfield95}-\cite{deCharms96}.
Then it is necessary for us to theoretically 
investigate high-dimensional, stochastic
differential equations (DEs) describing
the large-scale noisy neuron ensemble. 
In order to make our discussion concrete, let us consider
ensembles consisting of $N$-unit neurons, 
each of which is described by $K$-dimensional coupled DEs:
for example, $K$ =1, 2, 3 and 4 
for IF, FN, HR and HH neuron models, respectively.
Dynamics of such neuron ensembles, expressed by
$KN$-dimensional {\it stochastic} DEs, has been so far investigated
with the use of the two approaches: (i) direct simulations 
and (ii) analytical methods such as Fokker-Planck equation (FPE)
and the moment method.
Simulations have been made for large-scale
networks mostly consisting of IF neurons.
Since the CPU time to simulate networks by conventional methods 
is proportional to $N^2$, 
it is rather difficult to simulate realistic neuron clusters
in spite of recent computer development.  
In FPE dynamics of neuron ensembles is described by the 
population activity.
Although FPE is the powerful method formally applicable
to the case of arbitrary $K$ and $N$ \cite{FPE}, 
actual calculations have been made mostly for
$N=\infty$ ensembles of a $K=1$ model with the use of
the mean-field and/or diffusion approximations \cite{FPE2}.
Similar population density approaches have been recently developed
for a large-scale neuronal clusters \cite{Omurtag00,Haskell00}.
The moment method initiated by Rodriguez and Tuckwell (RT)
has been applied to single FN \cite{Rod96,Tuckwell98} 
and HH neurons \cite{Rod98,Rod00}.
When the moment method is applied to 
a single neuron model with $K$ variables, 
$K$-dimensional stochastic DEs 
are replaced by $(1/2)K (K+3)$-dimensional deterministic DEs.  
When the moment method is applied to $N$-unit neuron ensembles 
under consideration,
$K N$-dimensional stochastic DEs are replaced 
by $N_{eq}$-dimensional deterministic DEs
where $N_{eq}=(1/2)KN(KN+3)$ \cite{Rod96}.
For example, in the case of $K=2$ (FN model),
the number of equations is $N_{eq}=$ 230, 20 300 and 2 003 000 for
$N=$ 10, 100 and 1000, respectively.
In the case of $K=4$ (HH model),
we get $N_{eq}=$ 860, 80 600 and 8 006 000 for
$N=$ 10, 100 and 1000, respectively.
These figures are too large for us to make simulations
for realistic neuron clusters.
In their subsequent paper \cite{Rod98}, RT transplanted the result
of the moment method for HH neuron ensembles to FPE-type equation 
which has not been solved yet.

In a previous study 
(Ref.\cite{Hasegawa03a} is hereafter referred to as I),
the present author proposed 
a semi-analytical dynamical mean-field approximation (DMA), in which
equations of motions for means, variances and
covariances of {\it local} and {\it global} variables
were derived for $N$-unit FN neuron ensemble.
The original $2N$-dimensional stochastic DEs are replaced by 
eight-dimensional 
deterministic DEs: $N_{eq}=8$ is much smaller
than corresponding figures in the moment method
mentioned above.
DMA calculations in I on the spiking-time precision and
the synchronization in FN neuron ensembles
are in good agreement
with direct simulations. 
The feasibility of DMA has been demonstrated in I.

The purpose of the present paper is two folds.
The first purpose is to extend DMA of I to 
general neuron ensembles subject to white noises
described by $KN$-dimensional stochastic DEs, which will
be replaced by $K(K+2)$-dimensional deterministic DEs.
The second purpose of the present paper is to apply the
generalized DMA to an ensemble
of HH neurons, which is more realistic than FN neuron model
previously studied in I.
Since Hodgkin and Huxley proposed the HH model in 1952
\cite{Hodgkin52}, 
much studies have been intensively made on properties of the HH model.
Responses of a single, pairs and ensembles HH neurons
mostly to direct and sinusoidal currents have been investigated.
In recent years, responses of HH neurons to spike-train inputs 
have been studied \cite{Tanabe99}-\cite{Wang00b}.
The stochastic resonance (SR) of HH neurons 
for sinusoidal and spike inputs with
various kinds of added noises has been investigated
\cite{Lee98}-\cite{Hasegawa03b}.
These studies have shown that noise can play a constructive role 
in signal transmission against our conventional wisdom.
In most studies on SR, however, noises added to ensemble neurons
are considered to be independent of each other. 
Quite recently effects of spatially correlated noises
on SR have been investigated \cite{Liu00}, 
which shows that 
although common noises work to enhance
the synchronization in neuron ensembles,
they are not effective for SR, in contrast to
independent noises.
We will adopt in this study, 
spatially correlated white noises
in order to clarify respective effects
of common and independent noises on the
response of ensemble neurons.

The paper is organized as follows:
In Sec. II, we extend a DMA theory
to general neuron ensembles described by $KN$ stochastic DEs.
Our DMA theory is applied to HH neuron ensembles
in Sec. III. Some numerical results 
on HH neuron ensembles are presented in Sec. IV.
Conclusions and discussions are given in Sec. V.

\section{DMA for A General Neuron Ensemble}
\subsection{Equation of motions}

We assume an ensemble of $N$-unit neurons ($N \geq 2$),
each of which is described by $K$-dimensional
non-linear differential equations (DEs).
Dynamics of a given neuron ensemble is expressed by
\begin{eqnarray}
\frac{d v_{i}}{d t} &=& F^{(1)}(\{ u_{qi} \})
+\left( \frac{w}{N-1} \right)
\;\sum_{j (\neq i)} G(v_{j}(t))
+ K^{(e)}(t)+ \xi_i(t), \\
\frac{d u_{pi}}{d t} &=& F^{(p)}(\{ u_{qi} \}), 
\hspace{3cm}\mbox{($p=2$ to $K$)}
\end{eqnarray}
where $v_i=u_{pi}$ with $p=1$ denotes the membrane potential of
a neuron $i$ ($=1$ to $N$), $u_{pi}$ with $p=2$ to $K$ stands for
auxiliary variables and $F^{(p)}$ is functions of ($\{ u_{qi} \}$). 
The synaptic-coupling strength $w$ is assumed to be constant,
$G(v)=1/[1+{\rm exp}[-(v-\theta)/\epsilon]]$ is the sigmoid function
with the threshold $\theta$ and the width $\epsilon$ 
\cite{Note4,Note3}, 
and $K^{(e)}$ stands for an applied external
input whose explicit form will be given later [Eq. (56)].
The last term of Eq. (1) expresses the spatially
correlated white noises, $\xi_i(t)$,
given by
\begin{eqnarray}
<\xi_i(t)>&=&0, \\
<\xi_i(t)\:\xi_j(t')>
&=& [\beta_0^2 \;\delta_{ij}+\beta_1^2 \;
(1-\delta_{ij})]\:\delta(t-t') \nonumber \\
&=&(\beta_{C}^2+\beta_{I}^2 \; \delta_{ij})\:\delta(t-t'),
\end{eqnarray}
where $\beta_{C}=\beta_1$ and 
$\beta_{I}=\sqrt{\beta_0^2-\beta_1^2}$
denote the magnitudes of common and independent noises,
respectively, and the bracket $< \cdot >$ expresses 
the stochastic average \cite{Note2};
the case of $\beta_1=0$ ($\beta_1=\beta_0$) stands for
independent (common) noises only.

In order to derive DEs in DMA theory,
we first define the global variables for the ensemble 
by \cite{Hasegawa03a}
\begin{eqnarray}
U_{p}(t)&=&(1/N)\;\sum_{i} \;u_{pi}(t),
\end{eqnarray}
and their averages by
\begin{eqnarray}
\mu_{p}(t)&=&\mu_{u_p}(t)=<U_{p}(t)>.
\end{eqnarray}
Deviations from these averages of local variables are given by
\begin{eqnarray}
\delta u_{pi}(t)&=& u_{pi}(t)-\mu_{u_p}(t),
\end{eqnarray}
and those of global variables given by
\begin{eqnarray}
\delta U_p(t)&=& U_p(t)-\mu_{u_p}(t).
\end{eqnarray}

Next we define the variances and co-variances
between local variables given by 
(argument $t$ is neglected hereafter)
\begin{eqnarray}
\gamma_{p,q}&=&\gamma_{u_p,u_q}=
\frac{1}{N} \sum_i <\delta u_{pi} \: \delta u_{qi}>,
\end{eqnarray}
and those between global variables given by
\begin{eqnarray}
\rho_{p,q}&=&\rho_{u_p,u_q}=
<\delta U_p \: \delta U_q>, 
\end{eqnarray}
It is noted that $\gamma_{u_p,u_q}$ expresses fluctuations
in local variables while $\rho_{u_p,u_q}$
those in global variables.

We assume that the noise intensity is weak and that
the distribution function $p({\bf z})$ 
for $KN$-dimensional random variables of 
${\bf z}$ = ($\{ u_{pi} \}$) is given by
the Gaussian distribution concentrated near the
mean point of {\boldmath $\mu$}=$(\{\mu_{u_p} \})$ \cite{Note2}.
Numerical simulations
have shown that for weak noises, the distribution of $v(t)$ of the 
membrane potential of a single HH neuron nearly obeys
the Gaussian distribution,
although for strong noises, the distribution of $v(t)$
deviates from the Gaussian, 
taking a bimodal form \cite{Tanabe99}\cite{Tanabe01a}.
Similar behavior of the membrane-potential distribution 
has been reported also in a FN neuron model
\cite{Tuckwell98}\cite{Tanabe01}.
By using Eq. (7), we express Eqs. (1) and (2) in a Taylor expansion
of $\delta u_{pi}$ up to the fourth-order terms.
The average yields
DEs for the means of $d \mu_{u_p}/d t$ [Eq. (16)].
DEs of variances and covariances may be obtained 
by using the equations of motions of 
$\delta u_{pi}$.
For example, DE for $d \gamma_{u_p, u_q}/d t$ is given by
\begin{equation}
\frac{d \gamma_{u_p, u_q}}{d t} = \frac{1}{N} \sum_i
< \left(\frac{\partial \delta u_{pi}}{\partial t}\right) \;\delta u_{qi}
+\delta u_{pi} \;\left(\frac{\partial \delta u_{qi}}{\partial t}\right) >,
\end{equation} 
with 
\begin{eqnarray}
\frac{\partial \delta u_{pi}}{\partial t} 
&=& 
\sum_{q} F^{(u_p)}_{u_q} \;\delta u_{qi}
+ \frac{1}{2} \sum_q \sum_{r} F^{(u_p)}_{u_q u_r} 
(\delta u_{qi} \delta u_{ri} - \gamma_{u_q,u_r})  
+\frac{1}{6} \sum_q \sum_r \sum_s F^{(u_p)}_{u_q u_r u_s}\;
\delta u_{qi} \delta u_{ri} \delta u_{ti} \nonumber \\
&+& \delta_{p1} \;\{ \xi_i+\frac{1}{N-1} \sum_{k (\neq i)} 
[G_{u_1} \delta u_{1k} 
+ \frac{1}{2}  G_{u_1 u_1}(\delta u_{1k}^2-\gamma_{1,1})
+ \frac{1}{6} G_{u_1 u_1 u_1}\;\delta u_{1k}^3 ] + K^{(e)}  \},
\end{eqnarray}
where $q,\:r$ and $s$ run from 1 to $K$, 
$F^{(u_p)} = F^{(p)}$,
$F^{(u_p)}_{u_q} = \partial F^{(p)}/\partial u_{q}$,
$F^{(u_p)}_{u_q u_r} = \partial^2 F^{(p)}/ \partial u_{q}\partial u_{r}$ and
$F^{(u_p)}_{u_q u_r u_s} = \partial^3 F^{(p)}
/ \partial u_{q} \partial u_{r} \partial u_{s}$ 
evaluated at the
means of (\{$\mu_{u_p}$\}), and similar 
derivatives for $G$.
In the process of calculations of means, variances and covariances,
we have taken into account the fourth-order moment contributions
with the use of the Gaussian approximation, as given by
\begin{eqnarray}
<\delta u_{pi}\delta u_{qi}\delta u_{ri}\delta u_{si}>
&\simeq& <\delta u_{pi}\delta u_{qi}><\delta u_{ri}\delta u_{si}>
+<\delta u_{pi}\delta u_{ri}><\delta u_{qi}\delta u_{si}> \nonumber \\
&+&<\delta u_{pi}\delta u_{si}><\delta u_{qi}\delta u_{ri}>, \\
\frac{1}{N}
\sum_i <\delta u_{pi}\delta u_{qi}\delta u_{ri}\delta u_{si}>
&\simeq& \gamma_{u_p,u_q}\gamma_{u_r,u_s}+\gamma_{u_p,u_r}\gamma_{u_q,u_s}
+ \gamma_{u_p,u_s}\gamma_{u_q,u_r}, \\
\frac{1}{N^2} \sum_i \sum_j
<\delta u_{pi}\delta u_{qj}\delta u_{rj}\delta u_{sj}>
&\simeq& \rho_{u_p,u_q}\gamma_{u_r,u_s}+\rho_{u_p,u_r}\gamma_{u_q,u_s}
+ \rho_{u_p,u_s}\gamma_{u_q,u_r}, 
\end{eqnarray}
The importance of including the fourth-order term has
been pointed out by Tanabe and Pakdaman \cite{Tanabe01}
in the improved moment method for a noisy FN neuron.

After some manipulations,
we get DEs for means, variances and covariances
given by
(details being given in Appendix A of I):
\begin{eqnarray}
\frac{d \mu_{u_p}}{d t}&=&F^{(u_p)}
+\frac{1}{2} \sum_q \sum_r F^{(u_p)}_{u_q, u_r} \gamma_{u_q,u_r}
+ \delta_{p1} \;[ w\;U_0+K^{(e)} ], \\
\frac{d \gamma_{u_p,u_q}}{d t}&=& \sum_r 
[F^{(u_p)}_{u_r}\: \gamma_{u_q,u_r} + F^{(u_q)}_{r}\: \gamma_{u_p,u_r}]
+ \beta_0^2 \;\delta_{p1} \delta_{q1}
+ w\;U_1[\delta_{p1}\:\zeta_{u_q,u_1}
+ \delta_{q1}\:\zeta_{u_p,u_1}]
\nonumber \\
&&+\frac{1}{6} \sum_{r}\sum_{s}\sum_{w}
[ F^{(u_p)}_{u_r u_s u_w} 
(\gamma_{u_q,u_r}\gamma_{u_s,u_w} + \gamma_{u_q,u_s}\gamma_{u_r,u_w}
+ \gamma_{u_q,u_w}\gamma_{u_r,u_s}) \nonumber \\
&&+F^{(u_q)}_{u_r u_s u_w} 
(\gamma_{u_p,u_r}\gamma_{u_s,u_w} + \gamma_{u_p,u_s}\gamma_{u_r,u_w}
+ \gamma_{u_p,u_w}\gamma_{u_r,u_s}) ], \\
\frac{d \rho_{u_p,u_q}}{d t}&=& \sum_r 
[F^{(u_p)}_{u_r}\: \rho_{u_q,u_r} + F^{(u_q)}_{u_r}\: \rho_{u_p,u_r}]
+ [\frac{1}{N} \beta_0^2+\left( 1-\frac{1}{N} \right) \beta_1^2]
\;\delta_{p1}\delta_{q1}\nonumber \\
&&+ w\;U_1 [\delta_{p1}\:\rho_{u_q,u_1}
+ \delta_{q1}\:\rho_{u_p,u_1}] \nonumber \\
&&+\frac{1}{6} \sum_{r}\sum_{s}\sum_{t}
[ F^{(u_p)}_{u_r u_s u_t} 
(\rho_{u_q,u_r}\gamma_{u_s,u_t} + \rho_{u_q,u_s}\gamma_{u_r,u_t}
+ \rho_{u_q,u_t}\gamma_{u_r,u_s}) \nonumber \\
&&+F^{(u_q)}_{u_r u_s u_t} 
(\rho_{u_p,u_r}\gamma_{u_s,u_t} + \rho_{u_p,u_s}\gamma_{u_r,u_t}
+  \rho_{u_p,u_t}\gamma_{u_r,u_s}) ],
\end{eqnarray}
with
\begin{eqnarray}
\zeta_{u_p, u_q}
&=&\left( \frac{1}{N-1} \right)
(N \rho_{u_p, u_q}-\gamma_{u_p, u_q}), \\ 
U_0&=&\frac{1}{N} \sum_j <G(v_{j})>
= G + \frac{1}{2} G_{v v} \gamma_{v,v}, \\
U_1&=&G_{v} + \frac{1}{2} G_{v v v} \gamma_{v,v},
\end{eqnarray}
where $U_0$ expresses output spikes of the ensemble,
$v_j=u_{1j}$, and
arguments of $r$, $s$ and $w$ in the sums run from 1 to $K$.
The original $K\;N$-dimensional stochastic DEs are transformed to 
$N_{eq}$-dimensional deterministic DEs where 
$N_{eq}= K+K(K+1)=K(K+2)$.

\subsection{Property of DMA}

In previous Sec. IIA, DMA has been derived with the use of equations of motions
for moments.
It is, however, possible to alternatively derive DMA from the conventional
moment method with a reduction in numbers of variables, as was
shown in I for FN neuron ensembles.
In Appendix A, we present a derivation of DMA from the moment method
for a general neuron ensemble under consideration.

We should note that
the noise contribution is $\beta_0^2$ in Eq. (17)
while that is $[(1/N)\beta_0^2+(1-1/N)\beta_1^2]$ in Eq.(18).
When model parameters of $\beta_0$, $\beta_1$, $w$ and $N$ are varied,
the ratio of $\rho_{v,v}/\gamma_{v,v}$ changes.
In particular, in the case of $w=0$, we get
\begin{eqnarray}
\frac{\rho_{v,v}}{\gamma_{v,v}}
&=&\frac{1}{N}+\left( 1-\frac{1}{N} \right)
\:\left( \frac{\beta_1}{\beta_0} \right)^2, \\
&=& \frac{1}{N},
\hspace{2cm} \mbox{for $\beta_1=0$} \\
&=& 1.
\hspace{2.5cm} \mbox{for $\beta_1=\beta_0$}
\end{eqnarray}
Equation (23) agrees with the {\it central-limit theorem}
for independent noises
while Eq. (24) expresses the result for common noises.
On the other hand, in the opposite limit of $w \rightarrow \infty$,
we get $\rho_{v,v}/\gamma_{v,v} \rightarrow 1$.
The change in the ratio of 
$\rho_{v,v}/\gamma_{v,v}$ reflects on the 
the firing time distributions and the degree of synchronization
in neuron ensembles, as will be discussed in the followings.

\vspace{0.5cm}
\noindent
{\bf Firing Time Distributions}

The $n$th firing time of a given neuron $i$ in the
ensemble is defined as the time when
the membrane potential $v_i(t)$ 
crosses the threshold $\theta$ from below:
\begin{equation}
t_{oin}= 
\{ t \mid v_i(t) = \theta; \dot{v_i} >0 \}.
\end{equation} 
The distribution of firing times of $t_{oin}$ of a given neuron $i$
is given by \cite{Rod96}\cite{Hasegawa03a}
\begin{equation}
Z_{\ell}(t) \sim 
\phi\left( \frac{t-t^{*}_{o}}{\delta t_{o\ell}} \right) \;
\frac{d}{dt}\left( \frac{\mu_v}{\sigma_{\ell}} \right) \; 
\Theta(\dot{\mu_v}),
\end{equation} 
with the normal distribution function given by
\begin{equation}
\phi(x)=\frac{1}{\sqrt{2 \pi}} 
{\rm exp}\left( -\frac{x^2}{2} \right),
\end{equation} 
and 
\begin{equation}
\delta t_{o\ell}=\frac{\sigma_{\ell}}{\dot{\mu_v}},
\end{equation} 
where
$\sigma_{\ell}=\sqrt{\gamma_{v,v}}$
and $\dot{\mu_v}=d \mu_v/dt$ evaluated at $t=t_o^{*}$ where
$\mu_v(t_o^{*})=\theta$.
In the limit of vanishing $\beta$, Eq. (26) reduces to
\begin{equation}
Z_{\ell}(t)= \delta(t-t^{*}_{o}).
\end{equation} 

Similarly we may define the $m$th firing time relevant to 
the global variable $V(t)=(1/N) \sum_i v_i(t)$ as \cite{Hasegawa03a}
\begin{equation}
t_{gm}=\{ t \mid V(t) = \theta; \dot{V}(t) > 0 \}.
\end{equation} 
The distribution of firing times of $t_{gm}$ is given by
\begin{equation}
Z_{g}(t)= 
\phi \left( \frac{t-t^{*}_o}{\delta t_{og}} \right)\;
\frac{d}{dt} \left( \frac{\mu_1}{\sigma_g} \right) 
\; \Theta(\dot{\mu_v}),
\end{equation} 
with
\begin{equation}
\delta t_{og}=\frac{\sigma_g}{\dot{\mu_v}}
\end{equation} 
where $\sigma_g=\sqrt{\rho_{v,v}}$.
In particular, in the case of no couplings, we get
\begin{eqnarray}
\frac{\delta t_{og}}{\delta t_{o\ell}}
&=&\sqrt{\frac{1}{N}+(1-\frac{1}{N})
\:\left( \frac{\beta_1}{\beta_0}\right)^2}. 
\hspace{2cm} \mbox{(for $w=0$)}
\end{eqnarray}

\vspace{0.5cm}
\noindent
{\bf Synchronous Response}

The {\it synchronization ratio} is defined by
\cite{Hasegawa03a}
\begin{equation}
S(t)=\frac{(\rho_{v,v}/\gamma_{v,v}-1/N)}{(1-1/N)}
=\frac{\zeta_{v,v}}{\gamma_{v,v}},
\end{equation}
with
\begin{equation}
\zeta_{v,v}
=\left( \frac{1}{N-1}\right)(N\rho_{v,v}-\gamma_{v,v})
=\frac{1}{N(N-1)} 
\sum_i \sum_{j(\neq i)} <\delta v_{i}\;\delta v_{j}>,
\end{equation}
expressing the averaged covariance
for the variable of ($\{ \delta v_{i} \}$).
$S(t)$ changes as the model parameters of $\beta_0$, $\beta_1$,
$w$ and $N$ are varied.
It is easy to see from Eqs. (23) and (24) 
that $S=0$ (the asynchronous state) 
for $w =0 $ and $\beta_1 \ll \beta_0$, 
while $S=1$ (the completely synchronous state)
for $w \gg \beta_0^2$ or $\beta_1 = \beta_0$.
In particular, for $w=0$, we get
\begin{equation}
S(t)=\left( \frac{\beta_1}{\beta_0} \right)^2,
\hspace{2cm} \mbox{for $w=0$}
\end{equation}
which implies that the synchronization is induced
by common noises.

\section{DMA for HH Neuron Ensembles}
\subsection{Equation of Motions} 

For the HH neuron model ($K=4$), $F^{(p)}$ in Eq. (1)
is given by \cite{Hodgkin52}\cite{Hasegawa00}
\begin{eqnarray}
F^{(1)}=F^{(v)}(v_{i}, m_{i}, h_{i},n_{i})
&=& -\frac{1}{C}[g_{\rm Na} m_i^3 h_i (v_i - v_{\rm Na})
+ g_{\rm K} n_i^4 (v_i - v_{\rm K}) 
+ g_{\rm L} (v_i - v_{\rm L}) ], \\
F^{(p)}=F^{(u_p)}(v_{i}, u_{pi})
&=&- [a_{u_p}(v_i) + b_{u_p}(v_i)] \: u_{pi} + a_{u_p}(v_i),
\hspace{1cm} \mbox{($p=2$ to 4)}
\end{eqnarray}
In Eqs. (37) and (38) $u_{1i}=v_i$ expresses the membrane potential
of a neuron $i$, and 
$u_{2i}=m_i$, $u_{3i}=h_i$ and $u_{4i}=n_i$ denote
gate variables 
of Na and K channels for which 
$a_{u_p}(v)$ and $b_{u_p}(v)$ ($p=2$ to 4) are given by
\begin{equation}
a_m(v) = \frac{0.1 \: (v + 40)}{[1 - e^{-(v+40)/10}]},
\end{equation}
\begin{equation}
b_m(v) = 4 \: e^{-(v+65)/18},
\end{equation}
\begin{equation}
a_h(v) = 0.07 \: e^{-(v+65)/20},
\end{equation}
\begin{equation}
b_h(v) = \frac{1}{[1 + e^{-(v+35)/10}]}.
\end{equation}
\begin{equation}
a_n(v) = \frac{0.01 \: (v + 55)}{[1 - e^{-(v+55)/10}]},
\end{equation}
\begin{equation}
b_n(v) = 0.125 \: e^{-(v+65)/80}.
\end{equation}
In Eq. (37), 
the reversal potentials of
Na, K channels and leakage are $v_{\rm Na}=50$ mV, $v_{\rm K}=-77$ mV
and $v_{\rm L}=-54.5$ mV: 
the maximum values of corresponding conductances are
$g_{\rm Na}= 120$ ${\rm m S}/cm^2$,
$g_{\rm K}= 36$ ${\rm m S}/cm^2$ and
$g_{\rm L}= 0.3$ ${\rm m S}/cm^2$:
the capacitance of the membrane is $C=1$ $\mu {\rm F}/cm^2$.
From functional forms for $F^{(v)}$ and $F^{(u_p)}$
given by Eqs. (37)-(44), we get
$F_{v,v}=0, 
F^{(u_p)}_{u_q}= F^{(u_p)}_{u_p}\; \delta_{p\:q}, 
F^{(u_p)}_{v, u_q}= F^{(u_p)}_{v, u_p}\;\delta_{p\:q}$ and
$F^{(u_p)}_{u_p, u_q}= 0$.
Numbers of non-vanishing third-order derivatives are six for 
$F^{(v)}$ 
[$F^{(v)}_{vmm}$, $F^{(v)}_{vmh}$, $F^{(v)}_{vnn}$, 
$F^{(v)}_{mmm}$, $F^{(v)}_{nnn}$ and $F^{(v)}_{mmh}$]
and two for each $F^{(u_p)}$ $(p=2$ to 4)
[$F^{(u_p)}_{vvv}$ and $F^{(u_p)}_{vvu_p}$].

After some manipulations with Eqs. (16)-(18),
we get DEs for means, variances and covariances given by 
($p, q=2$ to 4)
\begin{eqnarray}
\frac{d \mu_{v}}{d t}&=&F^{(v)}
+ \frac{1}{2}\sum_{p=2}^4 \sum_{q=2}^4 
F^{(v)}_{u_p u_q}\: \gamma_{u_p, u_q}
+ \sum_{p=2}^4 F^{(v)}_{v u_p} \gamma_{v,u_p}
+ w \;U_0+K^{(e)}, \\
\frac{d \mu_{u_{p} }}{d t}&=& F^{(u_p)} 
+\frac{1}{2}\;F^{(u_p)}_{v, v} \gamma_{v, v}
+ F^{(u_p)}_{v, u_{p} } \gamma_{v, u_{p}},  \\
\frac{d \gamma_{v,v}}{d t}&=&2 [F^{(v)}_{v} \gamma_{v,v}
+ \sum_{p=2}^4 F^{(v)}_{u_p}\: \gamma_{v, u_p}]
+ \beta_0^2
+ 2\:w\;U_1\:\zeta_{v,v}
+X_{v,v}, \\
\frac{d \gamma_{v, u_{p}} }{d t}&=&
(F^{(v)}_{v}+ F^{(u_p)}_{u_{p} }) \:\gamma_{v, u_p}
+ \sum_{q=2}^4 F^{(v)}_{u_{q} } \gamma_{u_{q}, u_{p} }
+ F^{(u_p)}_{v } \gamma_{v, v} 
+w\;\zeta_{v, u_{p}} 
+X_{v,u_p},  \\
\frac{d \gamma_{u_{p}, u_{q} }}{d t}&=&
(F^{(u_p)}_{u_{p} }+F^{(u_q)}_{u_{q} } )
\gamma_{u_{p}, u_{q} }
+F^{(u_p)}_{v} \gamma_{v, u_q}+F^{(u_q)}_{v} \gamma_{v, u_p}+X_{u_p, u_q},  \\
\frac{d \rho_{v,v}}{d t}&=&2 [F^{(v)}_{v} \rho_{v,v}
+ \sum_{p=2}^4 F^{(v)}_{u_p}\: \rho_{v, u_p}]
+ [\frac{1}{N}\:\beta_0^2+(1-\frac{1}{N})\beta_1^2] 
+ 2\:w\;U_1\;\rho_{v,v}+Y_{v,v}, \\
\frac{d \rho_{v, u_{p}} }{d t}&=&
(F^{(v)}_{v}+ F^{(u_p)}_{u_{p} } )\:\rho_{v, u_p}
+ \sum_{q=2}^4 F^{(v)}_{u_{q} } \rho_{u_{q}, u_{p} } 
+ F^{(u_p)}_{v } \rho_{v, v} 
+w\:U_1\:\rho_{v, u_{p} } +Y_{v, u_p},  \\
\frac{d \rho_{u_{p}, u_{q} }}{d t}&=&
(F^{(u_p)}_{u_{p} }+F^{(u_q)}_{u_{q} } )
\rho_{u_{p}, u_{q} }
+F^{(u_p)}_{v} \rho_{v, u_q}+F^{(u_q)}_{v} \rho_{v, u_p} + Y_{u_p, u_q},  
\end{eqnarray}
with
\begin{eqnarray}
\zeta_{u_p, u_q}&=&\left( \frac{1}{N-1} \right)
(N \rho_{u_p, u_q}-\gamma_{u_p, u_q}), \\
U_0&=&\frac{1}{N}\sum_j <G(v_j)>
=G+\frac{1}{2}G_{vv}\gamma_{v,v}, \\
U_1&=&G_{v}+\frac{1}{2}G_{vvv}\gamma_{v,v}, 
\end{eqnarray}
where $F^{(v)}$, $F^{(v)}_v = \partial F^{(v)}/\partial v$ {\it et al.}
evaluated at means of $(\mu_v, \mu_m, \mu_h, \mu_n)$.
In Eqs. (45)-(52), $X_{v,v}$ and $Y_{v,v}$ {\it et al.} 
denote the contributions from
the fourth-order terms, whose explicit expressions are given by
Eqs. (B1)-(B6)
in Appendix B becuase they are rather lengthy.
Although calculations of the fourth-order terms are rather
tedious, they play important roles in stabilizing DEs.
This is numerically demonstrated in Appendix B for the
case of $N=1$.

The original $4N$-dimensional stochastic DEs given by Eqs. (37) and (38) 
are transformed to
24-dimensional deterministic DEs given by Eqs. (45)-(52) 
with Eqs. (B1)-(B6):
four means, ($\mu_v$, $\mu_m$, $\mu_h$, $\mu_n$),
ten moments for local variables
($\gamma_{v,v}$, $\gamma_{m,m}$, $\gamma_{h,h}$, $\gamma_{n,n}$,
$\gamma_{v,m}$, $\gamma_{v,h}$, $\gamma_{v,n}$,
$\gamma_{m,h}$, $\gamma_{h,n}$, $\gamma_{m,n}$), and
ten moments for global variables
($\rho_{v,v}$, $\rho_{m,m}$, $\rho_{h,h}$, $\rho_{n,n}$,
$\rho_{v,m}$, $\rho_{v,h}$, $\rho_{v,n}$,
$\rho_{m,h}$, $\rho_{h,n}$, $\rho_{m,n}$). 

In this subsection,
DMA for HH model has been obtained 
by the method of equations of motions of means, variances and covariances
of local and global variables.
We may, however, derive it from the moment method, as mentioned before.
In Appendix C, DEs in the moment method 
are presented for HH model.

We expect that our DMA equations given by Eqs.(45)-(52) and
(B1)-(B6) may show much variety depending 
on model parameters such
as the strength of white noise ($\beta_0$, $\beta_1$), 
couplings ($w$) and
the ensemble size ($N$).
In the next Sec. IV, we will present some numerical
DMA calculations, which are compared with simulation results. 
DMA equations have been solved by 
the fourth-order Runge-Kutta method
with a time step of 0.01 ms for the initial conditions of
$\mu_v=-65.0$, $\mu_m=0.0528$, $\mu_h=0.597$, $\mu_n=0.317$,
and $\gamma_{u_p, u_q}=\rho_{u_p, u_q}=0$
($u_p$, $u_q$=$v$, $m$, $h$ and $n$).
Direct simulations have been performed by solving
$4 N$-dimensional DEs given by  Eqs.(37) and (38) 
by using also
the fourth-order Runge-Kutta method with
a time step of 0.01.
Simulation results are the average of 100 trials otherwise noticed.

\section{Calculated Results of HH neuron ensembles}

\subsection{Firing time distribution}

In the present study, we pay our attention to
the response of the HH neuron ensembles
to a single spike input
applied to all neurons in the ensemble, given by
\begin{equation}
K^{(e)}(t)=\left( \frac{I_{i} }{C} \right) \;\alpha(t - t_{i}),
\end{equation}
with the alpha function:
\begin{equation}
\alpha(t)=\left( \frac{t}{\tau_s} \right) 
\;e^{(1 - t/\tau_s)} \;\Theta(t),
\end{equation}
where $\Theta(x)=1$ for $x \geq 0$ and 0 otherwise,  
$I_{i}$ stands for the magnitude of an input spike, 
$C$ the membrane capacitance [Eq. (37)],
$t_{i}$  the input time of a spike, and
$\tau_s$ (=1 ms) the time constant of synapses.
We get the critical magnitude of $I_{ic}$ = 3.62 $\mu$A/${\rm cm}^2$, 
below which firings of neuron cannot take place 
without noises ($\beta_0=\beta_1=0$). 
We have adopted the value of $I_{i}$ =5 $\mu$A/${\rm cm}^2$
for a study of the response to a supra-threshold input.
We express the coupling constant $w$ by $w=J/C$
with $J$ in units of $\mu$A/${\rm cm}^2$.
The time, voltage, current and noise intensity are hereafter expressed
in units of ms, mV, $\mu$A/${\rm cm}^2$ and V/s, respectively,
though they are sometimes omitted for a simplicity of
our explanation.
We have adopted parameters of  
$\theta=0$ mV and $\epsilon= 10$ mV in the sigmoid function $G(v)$
such that output $U_0$ is similar to the result given by
the alpha function [see Fig. 1(a)].
Adopted parameter values of $\beta_0$, $\beta_1$, $J$
and $N$ will be explained shortly.

Figures 1(a), 1(b) and 1(c) show
the time courses of $\mu_v$, $\sigma_{\ell}\:(=\sqrt{\gamma_{v,v}})$ 
and $\sigma_{g}\:(=\sqrt{\rho_{v,v}})$, respectively, 
when a single spike is applied at $t=100$ ms.
Solid and dashed curves express 
the results of DMA and direct simulations, respectively,
which are calculated with parameters of
$\beta_0=0.1$, $\beta_1=0$, $J=0$ and $N=100$.
States of neurons in an ensemble when an input spike
is injected at $t=100$ ms, are randomized because
noises have been already added since $t=0$.
We note that $\mu_v$ obtained by DMA
is in very good agreement with that obtained by simulations
as shown in Fig. 1(a), where
an external input of $K^{(e)}(t)$ and
an output of $U_0(t)$ are also plotted.
Figures 1(b) and 1(c) show that $\sigma_{\ell}$
and $\sigma_{g}$ calculated by DMA are again in
good agreement with those of simulations.
We note that the relation given by Eq.(22):
$\sigma_{g}=\sigma_{\ell}/N$ valid for $w=J/C=0$, 
is supported by our calculations.

Figure 2(a) shows $Z_{\ell}$, the firing probability
of local variable, which is calculated for $\beta_0=0.1$,
$\beta_1=0$, $J=0$ and $N=100$. 
Firings occur at $t \sim 103.6$ ms with a delay of about 3.6 ms.
Fluctuations of firing times of local variable, $\delta t_{o\ell}$, 
are 0.066 ms in DMA while it is 0.069 ms in simulations
which is the root-mean-square (RMS) value of firing times
defined by Eq. (25).
In contrast, Fig. 2(b) shows $Z_g$, the firing probability
of global variables.
Fluctuations of firing times of global variable $\delta t_{og}$ are 
0.0066 ms in DMA and it is 0.0083 ms in simulations, respectively.
We note that $\delta t_{og}$ is much smaller than $\delta t_{o\ell}$
[Eq. (33)].

\vspace{0.5cm}

\noindent
{\bf Noise-strength  dependence}

When the noise strength is increased,
the distribution of membrane potentials is widen
and fluctuations of firing times are increased,
as was discussed in Sec. IIB.
Filled squares in Fig. 3(a) show the $\beta_0$ dependence of
$\delta t_{o\ell}$ obtained by DMA theory
with $\beta_1=0$, $J=0$ and $N=100$, while
open squares express the RMS value of firing times
obtained by simulations.
The agreement between the two methods is in fairly good
for $\beta_0 < 0.1$ but becomes worse for $\beta > 0.1$.
In contrast, filled circles in Fig. 3(a) show 
the $\beta_0$ dependence of
$\delta t_{og}$ relevant to the global variable 
obtained by DMA theory
and open circles
stand for RMS values of firing times in simulations.
We note that $\delta t_{og}$ is much smaller
than $\delta t_{o\ell}$ 
because $\delta t_{og}=\delta t_{o\ell}/\sqrt{N}$ [Eq. (33)].

As $\beta_1$ is increased for a fixed $\beta_0$, the contribution from
common noises increases while that from independent noises
decreases 
($\beta_{C}=\beta_1$, 
$\beta_{I}=\sqrt{\beta_0^2-\beta_1^2}$).
The $\beta_1$ dependence of firing-time fluctuations
is shown in Fig. 3(b).
Filled squares and circles denote the results
of $t_{o\ell}$ and $t_{og}$, respectively, obtained by DMA,
and open squares and circles those by simulations.
Figure 3(b) shows that $\delta t_{og}$ is almost linearly
increased as $\beta_1$ is increased, 
while $\delta t_{o\ell}$ remains constant.
In the limit of $\beta_1 = \beta_0=0.1$, for which 
only common noises are applied 
($\beta_{C}=0.1$ and $\beta_{I}=0$),
we get $\delta t_{og}=\delta t_{o\ell}$, which
shows that common noises do not work
to reduce global fluctuations.
 
\vspace{0.5cm}

\noindent
{\bf Ensemble-size dependence}

Filled squares in Fig. 4(a) show the $N$ dependence of 
$\delta t_{o\ell}$ relevant to local fluctuations
for $\beta_0=0.1$, $\beta_1=0$ and $J=0$,
obtained by DMA theory, while
open squares express that obtained by simulations.
We note that $\delta t_{o\ell}$
is independent of $N$ because of no couplings ($J=0$).
In contrast,
$\delta t_{og}$ relevant to global fluctuations
inversely decreases 
when the size $N$ is increased, 
as shown by filled and open circles which are
obtained by DMA theory and simulations, respectively.
The relation: $\delta t_{og} \propto (1/\sqrt{N})$,
holds as given by Eq. (33) for $\beta_1=0$.
Figure 4(b) shows a similar plot for a finite value
of $\beta_1=0.05$ with $\beta_0=0.1$, $J=0$ and $N=100$.
In the limit of $N \rightarrow \infty$,
the ratio of $\delta t_{og}/\delta t_{o\ell}$ approaches
a finite value of $\beta_1/\beta_0=0.5$ [Eq. (33)].

\vspace{0.5cm}

\noindent
{\bf Coupling-strength dependence}

So far we have neglected the coupling of $J$,
which is now introduced.
Filled squares in Fig. 5(a) show the $J$ dependence 
of $\delta t_{o\ell}$ calculated by DMA theory 
for $\beta_0=0.1$, $\beta_1=0$ and $N=100$, 
while open squares that obtained by simulations. 
Filled and open circles express
$\delta t_{og}$
in the DMA theory and simulations, respectively.
We note that $\delta t_{o\ell}$ is
much reduced as $J$ is increased although there is little change 
in $\delta t_{og}$.
Figure 5(b) shows a similar plot of the $J$ dependence of
firing time accuracy for finite $\beta_1=0.05$ with 
$\beta_0=0.1$ and $N=100$.
Again a reduction in $\delta t_{o\ell}$ as increasing $J$
is more significant than that of $\delta_{og}$.

\subsection{Synchronization ratio}

One of important effects of the couplings is to yield
synchronous firings in ensemble neurons.
Figures 6(a) and 6(b) show
the time course of the synchronization ratio $S(t)$ 
for $J=100$ and 200 $\mu$A/${\rm cm}^2$, respectively,
with $\beta_0=0.1$, $\beta_1=0$ and $N=100$:
solid and dashed curves denote
the results of DMA and simulations, respectively. 
Fairly large fluctuations in simulation results are due to
a lack of trial number of one hundred, which
is a limit of our computer facility. 
A comparison between Figs. 6(a) and 6(b) shows that
$S(t)$ is increased as $J$ is increased:
the maximum value of $S(t)$ in Fig. 6(b) is $S_{max}=0.019$
which is larger than $S_{max}=0.007$ in Fig. 6(b).
Figure 6(c) shows the time course of $S(t)$ for a finite
$\beta_1=0.05$ with $\beta_0=0.1$, $J=100$ and $N=100$.
A significant increase in $S$ is realized 
at $100 \siml t \siml 120$ ms which is induced by an applied spike
[note the difference in vertical scales of Figs. 6(a), 6(b) and 6(c)].
We note a fairly large value of $S=0.25$ 
even without an applied input spike 
at $t \siml 100$ or $t \simg 120$.
This expresses the synchronization
among the membrane potentials of ensemble neurons
induced by added noises although they do not induce firings.
In order to distinguish the synchronization with firings from
that without firings, we define the firing-induced
synchronization ratio, $S'(t)$,  given by
\begin{equation}
S'(t) = S(t) - S_b,
\end{equation}
where $S_b=(\beta_1/\beta_0)^2$ denotes the {\it background} 
synchronization induced by noises only [Eq. (36)].
We get $S_{max}=0.369$, $S'_{max}=0.119$ and $S_b=0.25$ in Fig. 6(c).
From a comparison of Fig. 6(c) with Fig. 6(a),
we note that $S'(t)$ is also much increased by common noises.

An increase in $S(t)$ by an increase of $J$
is clearly shown in Fig. 7(a), where the maximum
of $S(t)$ ($S_{max}$) is plotted as
a function of $J$.
A disagreement between results of DMA and simulations
for $J < 50$ is due to fluctuations in simulations
because of insufficient trial number as mentioned above.
The dependence of $S_{max}$ on the size $N$
is shown in Fig. 7(b) where $\beta_0=0.1$,
$\beta_1=0$ and $J=100$.
$S_{max}$ is decreased as increasing $N$.
Figure 7(c) expresses the $\beta_0$ dependence of $S_{max}$
for $\beta_1=0.05$, $J=100$ and $N=100$.
At $\beta_0=\beta_1=0.05$, we get
$S_{max}=1$, which is decreased as increasing $\beta_0$.
Filled squares in Fig. 7( c) denote $S'_{max}$, which shows the
maximum around $\beta_0 \sim 0.08$.
In contrast, Fig. 7(d) show the $\beta_1$
dependence of $S_{max}$ for $\beta_0=0.1$, $J=100$ and $N=100$.
$S_{max}$ is increased as increasing $\beta_1$,
and approaches unity as $\beta_1 \rightarrow \beta_0$ (= 0.1).
We note that $S'_{max}$ has the maximum at $\beta_1 \sim 0.07$.

\section{Conclusion and Discussion}

In previous Sec. IV, we have reported DMA calculations
for a single spike input to HH neuron ensembles.
DMA calculations and simulations have shown that

(a) $\delta t_{o\ell}$ increases as increasing $\beta_0$, or decreasing $J$,
independently of $\beta_1$ and $N$,

(b) $\delta t_{og}$ increases as increasing $\beta_0$ or $\beta_1$, or
decreasing $N$, independently $J$, and

(c) $S_{max}$ increases as increasing $\beta_1$ or $J$, or decreasing $\beta_0$
or $N$.

In order to understand these behaviors,
we have tried to obtain phenomenological, analytical expressions for
$\delta t_{o\ell}$, $\delta t_{og}$ and $S_{max}$
as functions of $\beta_0$, $\beta_1$, $J$ and $N$.
For small $J$,
we express $\gamma_{v,v}$ and $\rho_{v,v}$ 
in power series of $J$ at $t=t_o^*$
where neurons fire, given by (see Appendix E of I, \cite{Note4})
\begin{eqnarray}
\gamma _{v,v}  &\propto& \beta _0^2 
\left[ {1 - \left( {a_1 J + a_2 J^2 +\cdot\cdot } \right)} \right], \\
\rho _{v,v}  &\propto& \beta _0 ^2 
\left[ {\frac{1}{N} + \left( {1 - \frac{1}{N}} \right)
\left( {\frac{{\beta _1 }}{{\beta _0 }}} \right)^2 } \right].
\end{eqnarray}
It is noted that in the limit of $J=0$ ($w=0$), Eqs. (59) and (60)
reduce to Eq. (22).
Substituting Eqs. (59) and (60) to Eqs. (28) and (32), we get
\begin{eqnarray}
\delta t_{o\ell}
&\propto&\beta_0 \left[1 
- \frac{1}{2}\left( {a_1 J + a_2 J^2 + \cdot \cdot } \right) \right], \\
\delta t_{og}
&\propto& \beta_0 \left[ {\frac{1}{N} + \left( {1 - \frac{1}{N}} \right)
\left( {\frac{{\beta _1 }}{{\beta _0 }}} \right)^2 } \right]^{1/2}. 
\end{eqnarray}
Equations (61) and (62) may explain the behavior of 
$\delta t_{o\ell}$ and $\delta t_{og}$ in
items (a) and (b) mentioned above.

Next we will obtain the analytical expression for 
$S_{max}$.
For small $J$, we get 
$\gamma_{v,v}$ and $\rho_{v,v}$ in power series of $J$
at $t=t_o^{(m)}$
where $S(t)$ takes the maximum value, 
given by (see Appendix E of I, \cite{Note4})
\begin{eqnarray}
\gamma _{v,v}  &\propto& \beta _0^2 
\left[ {1 - \left( {b_1 J + b_2 J^2 + \cdot\cdot } \right)} \right], \\
\rho _{v,v}  &\propto& \beta _0 ^2 
\left[ {\frac{1}{N} + \left( {1 - \frac{1}{N}} \right)
\left( {\frac{{\beta _1 }}{{\beta _0 }}} \right)^2 } \right].
\end{eqnarray}
Substituting Eqs. (63) and (64) to Eq. (34), we get 
\begin{eqnarray}
S_{\max} &=& \left( {\frac{{\beta _1 }}{{\beta _0 }}} \right)^2  
+\left[ {\frac{1}{N} + \left( {1 - \frac{1}{N}} \right)
\left( {\frac{{\beta _1 }}{{\beta _0 }}} \right)^2 } \right]
\left[ {b_1 J  + (b_2+b_1^2) J^2 + \cdot\cdot } \right].
\end{eqnarray}
Equation (65) is consistent with the item (c):
$S_{max}$ increases as increasing $\beta_1$
and $J$, or decreasing $\beta_0$ and $N$.
In the case of $\beta_1=0$, Eq. (65) shows that $S_{max}$ 
is independent of $\beta_0$, which is supported in DMA calculation
and simulations (not shown).
Expressions given by Eqs. (61), (62) and (65)
are useful in a phenomenological sense.
In principle, expressions as given by Eqs. (59) and (60)
may be derived from DMA equations given by Eqs. (45)-(52)
although we have not unfortunately succeeded in getting them
because of their complexity.  

Numerical calculations in Sec. IV have been made for
the response to a single spike input. 
DMA is, however, applicable to arbitrary inputs. 
This will be demonstrated by adding spike trains 
to HH neuron ensembles, given by
\begin{equation}
K^{(e)}(t)=(\frac{I_{i} }{C}) \sum_{n}\;\alpha(t - t_{in}),
\end{equation}
where $t_{in}$ expresses the $n$th input time.
Figures 8(a) and 8(b) show the time courses of $\mu_v$
and $\sigma_{\ell}$ (=$\sqrt{\gamma_{v,v}}$), respectively,
for Poisson spike trains
with the average interspike interval (ISI) of 25 ms;
solid and dashed curves express results
DMA and simulations, respectively,
for $\beta_0=0.1$, $\beta_1=0$, $J=0$ and $N=100$.
The time course of $\mu_v$ of DMA is in good agreement with that
of simulations. A comparison between the input $K^{(e)}$ 
and output $U_0$ shows 
that when ISI of input is shorter than about 10 ms, HH neurons
cannot respond because of the refractory period.
Figures 8(b) shows that $\sigma_{\ell}$ 
of DMA is also in good agreement with that of simulations.


To summarize,
the DMA theory previously proposed for FN neuron ensemble in I, has 
been generalized to an ensemble described by
$KN$-dimensional stochastic DEs, 
which has been be replaced by $K(K+2)$-dimensional 
deterministic DEs expressed by means and second-order moments: contributions
from the fourth-order moments are taken in account
by the Gaussian decoupling approximation.
DMA has been applied to HH neuron ensembles, for which we get
24-dimensional deterministic DEs.
We have studied effects of noise, the coupling strength and the ensemble size
on the firing time precision and the firing synchronization
for single-spike inputs, obtaining the following results:

(i) the firing-time accuracy of the order of one-tenth ms is
possible in a large-scale HH neuron ensemble, even without couplings,

(ii) the spike transmission is improved with the synchronous response 
by increasing the coupling strength, and

(iii) the synchronization is increased by common noises 
but decreased by independent noises.

The item (i) and (ii) are consistent with the SR results
in HH neuron ensembles \cite{Pei96a}-\cite{Hasegawa03b}. 
Although they are quite similar to the case of FN model discussed
in I, their {\it quantitative} discussions are possible
with the use of the realistic HH model. 
The item (iii) agrees with the result
of Ref. \cite{Liu00} for SR in HH neuron ensembles subject of
common and independent noises. 

Our calculations have demonstrated the feasibility of DMA,
whose advantages may be summarized as follows:

(1) because of the semi-analytical nature of DMA, some results
may be derived without numerical calculations,

(2) DMA is free from the weak-coupling constraint
although it assumes weak noises,

(3) a tractable small number of DEs makes calculations feasible
for large-scale neuron ensembles with a fairly short computational time,

(4) DMA may be applicable to ensembles 
with fluctuations not only due to noises but also
due to some inhomogeneities in model parameters, and 

(5) DMA can be applied to more general stochastic systems besides 
neuron models.

As for the item (3), we may point out that
for example, the CPU time 
of DMA calculations for a 200 ms time course
of a $N=100$ HH neuron ensemble with the use of 1.8 GHz PC 
is 2 s, which is about 2500 times faster than
the CPU time of 85 min ($\sim$ 5000 s) 
for direct simulations with 100 trials. 
It is necessary to stress the importance
of the fourth-order contributions
in stabilizing solutions of DMA,
which is numerically demonstrated in Appendix B.
Although expressions for the fourth-order contributions
are lengthy, we are much benefited from them 
once they are derived and planted into 
computer programs \cite{Note1}. 
This paper is the first report on our DMA calculations of HH neuron
ensembles.
We are now under consideration to incorporate
the time delay in the coupling terms in Eq. (1),
with which the HH neuron ensemble may show the intrigue behavior like
chaos.
Such calculations will be reported in a separate paper.

\section*{Acknowledgments}
The author would like to express his sincere thanks to
Professor Hideo Nitta for critical reading of the manuscript.
This work is partly supported by
a Grant-in-Aid for Scientific Research from the Japanese 
Ministry of Education, Culture, Sports, Science and Technology.  

\newpage

\appendix
\section{Derivation of DMA from the moment method for general
neuron ensembles}

In the moment method, we define the means, variances
and covariances given by \cite{Rod96}
\begin{eqnarray}
m_{u_p}^i&=& <u_{pi}>, \\
C_{u_p,u_q}^{i,j} &=& <\Delta u_{pi} \Delta u_{qj}>, 
\end{eqnarray}
where $\Delta u_{pi}=u_{pi}-m_{u_p}^{i}$.
Assuming the weak couplings and adopting the Gaussian
decoupling approximations for the fourth-order moments, we get DEs for
general neuron ensembles described by Eqs. (1) and (2):
\begin{eqnarray}
\frac{d m_{u_p}^i}{d t}&=&F^{(u_{p})}
+\frac{1}{2}\sum_{q} \sum_{r} F^{(u_{p})}_{u_{qi}, u_{ri}}
\: C^{(i,i)}_{u_q, u_r} \nonumber \\
&&+ \delta_{p1}
[\left( \frac{w}{N-1}\right) \;\sum_{k(\neq i)}
(G+\frac{1}{2}G_{u_{1k}u_{1k}} C_{u_1,u_1}^{k,k})
+K^{(e)}],\\
\frac{d C^{i,j}_{u_p,u_q}}{d t}&=& \sum_r 
[F^{(u_p)}_{u_{ri}}\: C^{i,j}_{u_q,u_r} 
+ F^{(u_q)}_{u_{rj}}\: C^{i,j}_{u_p,u_r}]
+[\beta_0^2\;\delta_{ij}+\beta_1^2\;(1-\delta_{ij})]
\;\delta_{p1}\delta_{q1} \nonumber\\
&&+\delta_{p1}\left( \frac{w}{N-1} \right)
\sum_{k(\neq i)} G_{u_{1k}}\:C^{j,k}_{u_q,u_1}
+\delta_{q1} \left( \frac{w}{N-1} \right)
\sum_{k(\neq j)} G_{u_{1k}}\:C^{i,k}_{u_p,u_1} \nonumber \\
&&+\frac{1}{6} \sum_{r}\sum_{s}\sum_{w}
[ F^{(u_p)}_{u_{ri} u_{si} u_{wi}} 
(C_{u_r,u_s}^{i,i}C_{u_w,u_q}^{i,j} 
+ C_{u_r,u_w}^{i,i}C_{u_s,u_q}^{i,j}
+ C_{u_s,u_w}^{i,i}C_{u_r,u_q}^{i,j}) \nonumber \\
&&+F^{(u_q)}_{u_{rj} u_{sj} u_{wj}} 
(C_{u_r,u_s}^{j,j}C_{u_w,u_p}^{j,i} 
+ C_{u_r,u_w}^{j,j}C_{u_s,u_p}^{j,i}
+ C_{u_s,u_w}^{j,j}C_{u_r,u_p}^{j,i}) ],
\end{eqnarray}
where $F^{(u_p)}=F^{(p)}$,
$F^{(u_p)}_{u_{ri}}=\partial F^{(p)}/\partial u_{ri}$ and 
$F^{(p)}_{u_{ri}u_{si}u_{wi}}
=\partial^{(3)} F^{(p)}/\partial u_{ri}\partial u_{si}\partial u_{wi}$
evaluted for the means of $(\{ m^i_{u_p} \})$,
and the last term in Eq. (A4) denotes the fourth-order contribution.
The number of DEs is $N_{eq}=KN+(1/2)KN(KN+1)=(1/2)KN(KN+3)$.

In order to derive DMA from the moment method, 
we define the quantities given by
\begin{eqnarray}
\overline{\mu}_{\kappa}&=&\frac{1}{N} \sum_i m^i_{\kappa}, \\
\overline{\gamma}_{\kappa, \lambda}
&=& \frac{1}{N} \sum_i C^{i,i}_{\kappa, \lambda}
+ d_{\kappa, \lambda}, \\
\overline{\rho}_{\kappa, \lambda}
&=& \frac{1}{N^2} \sum_i \sum_j C^{i,j}_{\kappa, \lambda}, 
\hspace{1cm} \mbox{($\kappa, \lambda= u_{p},u_q$)}
\end{eqnarray}
where
\begin{eqnarray}
d_{\kappa, \lambda}&=&\frac{1}{N} 
\sum_i \delta m^i_{\kappa} \delta m^i_{\lambda}, \\
\delta m^i_{\kappa}&=&m^i_{\kappa}-\mu_{\kappa}.
\end{eqnarray}
We may show that 
Eqs. (A3) and (A4) with Eqs. (A3)-(A7) yield Eqs. (16)-(18):
$\overline{\mu}_{\kappa}=\mu_{\kappa}$,
$\overline{\gamma}_{\kappa, \lambda}=\gamma_{\kappa,\lambda}$ and
$\overline{\rho}_{\kappa, \lambda}=\rho_{\kappa,\lambda}$.
Then the moment method yields the same results as DMA 
as far as the averaged quantities are concerned
(see also Appendix B of I). 

\section{The fourth-order contributions in DMA 
for HH neuron ensembles}

The fourth-order contributions given by $X_{\kappa,\lambda}$
and $Y_{\kappa,\lambda}$ ($\kappa, \lambda=v, u_{p}$) 
in Eqs. (45)-(52) are expressed by 
\begin{eqnarray}
X_{v,v}&=& 
F^{(v)}_{vmm} (\gamma_{v,v} \gamma_{m,m}+2 \gamma_{v,m} \gamma_{v,m})
+F^{(v)}_{vmh} (\gamma_{v,v} \gamma_{m,h}+2 \gamma_{v,m} \gamma_{v,h}) 
\nonumber \\
&&+F^{(v)}_{vnn} (\gamma_{v,v} \gamma_{n,n}+2 \gamma_{v,n} \gamma_{v,n})
+F^{(v)}_{mmm} \gamma_{v,m}\gamma_{m,m}
+F^{(v)}_{nnn} \gamma_{v,n}\gamma_{n,n} \nonumber \\
&&+F^{(v)}_{mmh} (\gamma_{v,h} \gamma_{m,m}+2 \gamma_{v,m} \gamma_{m,h}), \\
X_{v,u_p}&=& \frac{1}{2}[
F^{(v)}_{vmm} (\gamma_{v,u_p} \gamma_{m,m}+2 \gamma_{u_p,m} \gamma_{v,m})
+F^{(v)}_{vmh} (\gamma_{v,u_p} \gamma_{m,h}
+\gamma_{u_p,m} \gamma_{v,h} +\gamma_{u_p,h} \gamma_{v,m}) 
\nonumber\\
&&+F^{(v)}_{vnn} (\gamma_{v,u_p} \gamma_{n,n}+2 \gamma_{u_p,n}\gamma_{v,n} )
+F^{(v)}_{mmm} \gamma_{u_p,m}\gamma_{m,m}
+F^{(v)}_{nnn} \gamma_{u_p,n}\gamma_{n,n} \nonumber \\
&&+F^{(v)}_{mmh} (\gamma_{u_p,h} \gamma_{m,m}+2 \gamma_{u_p,m} \gamma_{m,h})
+F^{(u_p)}_{vvv} \gamma_{v,v} \gamma_{v,v} \nonumber \\
&&+F^{(u_p)}_{vvu_p} (\gamma_{v,u_p}\gamma_{v,v}+2 \rho_{v,v}\gamma_{v,u_p}) ], 
\\
X_{u_p,u_q}&=& \frac{1}{2} [
F^{(u_p)}_{vvv}\gamma_{v,u_q}\gamma_{v,v}
+F^{(u_q)}_{vvv} \gamma_{v,u_p}\gamma_{v,v}
+(F^{(u_p)}_{vvu_p}+F^{(u_q)}_{vvu_q} )
(\gamma_{u_p,u_q}\gamma_{v,v}+2 \gamma_{v,u_q}\gamma_{v,u_p})],\\ 
Y_{v,v}&=& 
F^{(v)}_{vmm} (\rho_{v,v} \gamma_{m,m}+2 \rho_{v,m} \gamma_{v,m})
+F^{(v)}_{vmh} (\rho_{v,v} \gamma_{m,h}+\rho_{v,m} \gamma_{v,h}
+ \rho_{v,h} \gamma_{v,m}) \nonumber \\
&&+F^{(v)}_{vnn} (\rho_{v,v} \gamma_{n,n}+2 \rho_{v,n} \gamma_{v,n})
+F^{(v)}_{mmm} \rho_{v,m}\gamma_{m,m}
+F^{(v)}_{nnn} \rho_{v,n}\gamma_{n,n} \nonumber \\
&&+F^{(v)}_{mmh} (\rho_{v,h} \gamma_{m,m}+2 \rho_{v,m} \gamma_{m,h}), \\
Y_{v,u_p}&=& \frac{1}{2}[
F^{(v)}_{vmm} (\rho_{v,u_p} \gamma_{m,m}+2 \rho_{u_p,m} \gamma_{v,m})
+F^{(v)}_{vmh} (\rho_{v,u_p} \gamma_{m,h}
+\rho_{u_p,m} \gamma_{v,h} +\rho_{u_p,h} \gamma_{v,m}) 
\nonumber\\
&&+F^{(v)}_{vnn} (\rho_{v,u_p} \gamma_{n,n}+2 \rho_{u_p,n}\gamma_{v,n} )
+F^{(v)}_{mmm} \rho_{u_p,m}\gamma_{m,m}
+F^{(v)}_{nnn} \rho_{u_p,n}\gamma_{n,n} \nonumber \\
&&+F^{(v)}_{mmh} (\rho_{u_p,h} \gamma_{m,m}+2 \rho_{u_p,m} \gamma_{m,h})
+F^{(u_p)}_{vvv} \rho_{v,v} \gamma_{v,v} \nonumber \\
&&+F^{(u_p)}_{vvu_p} (\gamma_{v,u_p}\gamma_{v,v}+2 \rho_{v,v}\gamma_{v,u_p}) ], \\
Y_{u_p,u_q}&=& \frac{1}{2} [
F^{(u_p)}_{vvv} \rho_{v,u_q}\gamma_{v,v}
+F^{(u_p)}_{vvu_p}(\rho_{u_p,u_q}\gamma_{v,v}+2 \rho_{v,u_q}\gamma_{v,u_p}) 
+F^{(u_q)}_{vvv}\rho_{v,u_p} \gamma_{v,v} \nonumber\\
&&+F^{(u_q)}_{vvu_q}(\rho_{u_p,u_q}\gamma_{v,v}+2 \rho_{v,u_p}\gamma_{v,u_q}) ]
\end{eqnarray}
where
$F^{(v)}_{vmh}=\partial^{3} F^{(v)}/\partial v \partial m \partial h$
{\it et al.}
Although calculations and 
computer programming of fourth-order contributions 
given by Eqs. (B1)-(B6) are rather tedious,
they play important roles in stabilizing the solution of DEs \cite{Note1}.   

Here we demonstrate the importance of the fourth-order
contributions in the case of a single HH neuron ($N=1$)
for which $w=0$, $\gamma_{\kappa,\lambda}=\rho_{\kappa,\lambda}$ and 
$X_{\kappa,\lambda}=Y_{\kappa,\lambda }$
in Eqs. (45)-(52) and (B1)-(B6).
Figure 9(a) shows the time course of $\mu_v$ 
for $\beta_0=0.1$ and $\beta_1=0$
when the constant input of $I_{i}=10$ $\mu$A/${\rm cm}^2$ is
applied at $t=0$ ms.
The solid and dashed curves express the results
of DMA and the simulation (100 trials), respectively.
The dotted curve denote the result of DMA2
(the second-order DMA)
in which the fourth-order contributions are neglected
($X_{\kappa,\lambda}=Y_{\kappa,\lambda }=0$).
For $t < 60$ ms, all results seem to be in good agreement.
At $t \simg 60$, however, the solution of DMA2
becomes unstable and significantly deviates from those
of DMA and the simulation. 
From the time course of 
$\sigma_{\ell
}=\sqrt{\gamma_{v,v}}$ shown in Fig. 9(b),
we note that such deviation of DMA2 already starts from $t \sim 30$ ms.
The solution of DMA2 is stable at $\beta \leq 0.037$ 
for the constant current of $I_{i}=10$ $\mu$A/${\rm cm}^2$.

Figures 10(a) and 10(b) show the time courses of 
$\mu_v$ and $\sigma_{\ell}$
for $\beta_0=0.2$ and $\beta_1=0$ 
when we apply the periodic spike train input given by
Eq. (66) with $I_{i}=5$ $\mu$A/${\rm cm}^2$ and a constant ISI of 25 ms.
Figure 10(b) clearly shows that the result of DMA2 deviates
from those of DMA and simulations from the first spike input
and that the result of DMA2 diverges at the second spike input.
The solution of DMA2 is stable only at $\beta \leq 0.178$ 
for this periodic spike.


\section{The moment method for HH neuron ensembles}

We will derive DEs in the moment method for HH neuron ensembles,
defining the means, variances
and covariances given by \cite{Rod96}
\begin{eqnarray}
m_v^i&=&<v_i>, \\
m_{u_p}^i&=& <u_{pi}>, \\
C_{v,v}^{i,j}&=& <\Delta v_i \Delta v_j>, \\
C_{v, u_p}^{i,j} &=& <\Delta v_i \Delta u_{pi}>, \\
C_{u_p,u_q}^{i,j} &=& <\Delta u_{pi} \Delta u_{qj}>, 
\end{eqnarray}
where $\Delta v_i = v_i - m_v^i$ and 
$\Delta u_{pi}=u_{pi}-m_{u_p}^{i}$.
Adopting the weak-coupling approximation
and the Gaussian approximation for the fourth-order terms, 
we get DEs given by
\begin{eqnarray}
\frac{d m_{v}^i}{d t}&=&F^{(v_i)}
+\frac{1}{2}\sum_{p=2}^4 \sum_{q=2}^4 
F^{(v_i)}_{u_p, u_q}\: C_{u_p, u_q}^{i,i}
+ \sum_{p=2}^4 F^{(v_i)}_{v, u} C_{v,u_p}^{i,i} \nonumber \\
&&+ \left( \frac{w}{N-1} \right)
\;\sum_{k(\neq i)}(G^{(v_k)}
+\frac{1}{2}G^{(v_k)}_{vv} C_{v,v}^{k,k})
+K^{(e)}(t), \\
\frac{d m_{u_{p} }^i}{d t}&=& F^{(u_{pi})} 
+\frac{1}{2}F^{(u_{pi})}_{v, v} C_{v, v}^{i,i}
+ F^{(u_{pi})}_{v, u_{p} } C_{v, u_{p}}^{i,i},  \\
\frac{d C_{v,v}^{i,j}}{d t}&=&
2 F^{(v_i)}_{v} C_{v,v}^{i,j}
+ \sum_{p=2}^4 F^{(v_i)}_{u_p}\:(C_{v, u_p}^{i,j}+ C_{v, u_p}^{j,i} )
+[\beta_0^2\;\delta_{ij}+\beta_1^2\;(1-\delta_{ij})] \nonumber \\
&&+ \left( \frac{w}{N-1} \right) 
[\sum_{k(\neq i)} G^{(v_k)}_v\:C_{v,v}^{j,k} 
+ \sum_{k(\neq j)} G^{(v_k)}_v\:C_{v,v}^{i,k}]
+Z_{v,v}^{i,j}, \\
\frac{d C_{v, u_{p}}^{i,j} }{d t}&=&
(F^{(v_i)}_{v}+ F^{(u_{pi})}_{u_{p} }) \:C_{v, u_p}^{i,j}
+ \sum_{q=2}^4 F^{(v_i)}_{u_{q} } C_{u_{q}, u_{p} }^{i,j}
+ F^{(u_{pi})}_{v } C_{v, v}^{i,j} \nonumber \\
&&+ \left( \frac{w}{N-1} \right)
\sum_{k(\neq i)} G^{(v_k)}_v\:C_{v, u_{p} }^{k,j}
+Z_{v,u_p}^{i,j},  \\
\frac{d C_{u_{p}, u_{q} }^{i,j}}{d t}&=&
(F^{(u_{pi})}_{u_{p} } C_{u_{q}, u_{p} }^{i,j} 
+F^{(u_{qi})}_{u_{q} } C_{u_{p}, u_{q} }^{i,j} )
+F^{(u_{pi})}_{v} C_{v, u_q}^{i,j}
+F^{(u_{qi})}_{v} C_{v, u_p}^{j,i}+Z_{u_p,u_q}^{i,j},
\end{eqnarray}
with
\begin{eqnarray}
Z_{v,v}^{i,j}&=&\frac{1}{2}[ 
F^{(v)}_{vmm} (C^{i,j}_{v,v} C^{j,j}_{m,m}+2 C^{i,j}_{v,m} C^{j,j}_{v,m}
+C^{i,j}_{v,v} C^{i,i}_{m,m}+2 C^{j,i}_{v,m} C^{i,j}_{v,m}) \nonumber \\
&&+F^{(v)}_{vmh} (C^{i,j}_{v,v} C^{j,j}_{m,h}+C^{i,j}_{v,m} C^{j,j}_{v,h}
+ C^{i,j}_{v,h} C^{j,j}_{v,m}
+C^{i,j}_{v,v} C^{i,i}_{m,h}+C^{j,i}_{v,m} C^{i,i}_{v,h}
+ C^{j,i}_{v,h} C^{i,i}_{v,m}) \nonumber \\
&&+F^{(v)}_{vnn} (C^{i,j}_{v,v} C^{j,j}_{n,n}
+2 C^{i,j}_{v,n} C^{j,j}_{v,n}
+C^{i,j}_{v,v} C^{i,i}_{n,n}+2 C^{j,i}_{v,n} C^{i,i}_{v,n}) \nonumber\\
&&+F^{(v)}_{mmm} (C^{i,j}_{v,m}C^{j,j}_{m,m}+C^{j,i}_{v,m}C^{i,i}_{m,m})
+F^{(v)}_{nnn} (C^{i,j}_{v,n}C^{j,j}_{n,n}+ C^{j,i}_{v,n}C^{i,i}_{n,n})
\nonumber \\
&&+F^{(v)}_{mmh} (C^{i,j}_{v,h} C^{j,j}_{m,m}+2C^{i,j}_{v,m} C^{j,j}_{m,h}
+C^{j,i}_{v,h} C^{i,i}_{m,m}+2C^{j,i}_{v,m} C^{i,i}_{m,h})], \\
Z_{v,u_p}^{i,j}&=& \frac{1}{2}[
F^{(v)}_{vmm} (C^{i,j}_{v,u_p} C^{i,i}_{m,m}+2 C^{i,j}_{m,u_p} C^{i,i}_{v,m})
+F^{(v)}_{vmh} (C^{i,j}_{v,u_p} C^{i,i}_{m,h}
+C^{i,j}_{m,u_p} C^{i,i}_{v,h} +C^{i,j}_{h,u_p} C^{i,i}_{v,m}) 
\nonumber\\
&&+F^{(v)}_{vnn} (C^{i,j}_{v,u_p} C^{i,i}_{n,n}+2C^{i,j}_{n,u_p}C^{i,i}_{v,n})
+F^{(v)}_{mmm} C^{i,j}_{m,u_p} C^{i,i}_{m,m}
+F^{(v)}_{nnn} C^{i,j}_{n,u_p} C^{i,i}_{n,n} \nonumber \\
&&+F^{(v)}_{mmh} (C^{i,j}_{h,u_p} C^{i,i}_{m,m}+2 C^{i,j}_{m,u_p}C^{i,i}_{m,h})
+F^{(u_p)}_{vvv} C^{i,j}_{v,v} C^{j,j}_{v,v} \nonumber \\
&&+F^{(u_p)}_{vvu_p} (C^{i,j}_{v,u_p}C^{j,j}_{v,v}
+2 C^{i,j}_{v,v} C^{j,j}_{v,u_p}) ], \\
Z_{u_p,u_q}^{i,j}&=& \frac{1}{2} [
F^{(u_p)}_{vvv} C^{i,j}_{v,u_q} V^{i,i}_{v,v}
+F^{(u_p)}_{vvu_p}(C^{i,j}_{u_p,u_q} C^{i,i}_{v,v}
+2 C^{i,j}_{v,u_q} C^{i,i}_{v,u_p}) 
+F^{(u_q)}_{vvv} C^{i,j}_{u_p,v} C^{j,j}_{v,v} \nonumber\\
&&+F^{(u_q)}_{vvu_q}(C^{i,j}_{u_p,u_q} C^{j,j}_{v,v}
+2 C^{i,j}_{u_p,v} C^{j,j}_{v,u_q}) ], 
\end{eqnarray}
where $F^{(v_i)}$, $F^{(v_i)}_v=\partial F/\partial v_i$ and
$F^{(v_i)}_{v,u_p}=\partial^2 F/\partial v_i \partial u_{pi}$
evaluated for the averages of $(m^i_v, \{ m^i_{u_p}\})$,
and similar derivatives for $F^{(u_{pi})}$ and $G^{(v_k)}$.
Equations given by Eqs. (C6)-(C13) denote the result of
the fourth-order moment method.
The second-order moment method was applied to a single
HH neuron by RT \cite{Rod98}\cite{Rod00}, 
whose result is given by Eqs. (C6)-(C10)
when we set $i=j=1$, $w=0$, $\beta_1=0$ 
and $Z_{v,v}=Z_{v,u_p}=Z_{u_p,u_q}=0$.
Equations (C6)-(C13) lead to DMA equation given by
Eqs. (45)-(52) and (B1)-(B6) 
if we adopt the relations as given by Eqs. (A5)-(A7).
In particular, in the case of $N=1$, Eqs. (C6)-(C13)
are identical with Eqs. (45)-(52) and (B1)-(B6) of DMA if
we read $m_{\kappa}^1=\mu_{\kappa}$,
$C_{\kappa,\lambda}^{1,1}
=\gamma_{\kappa,\lambda}
=\rho_{\kappa,\lambda}$, and
$Z_{\kappa,\lambda}^{1,1}=X_{\kappa,\lambda}
=Y_{\kappa,\lambda}$.



\begin{figure}
\caption{
Time courses of 
(a) $\mu_v$, (b) $\sigma_{\ell}$ ($=\surd{\gamma_{v,v}}$) and
(c) $\sigma_{g}$ ($=\surd{\rho_{v,v}})$,
for $\beta_0=0.1$, $\beta_1=0$, $J=0$ and $N=100$,
$K^{(e)}$ and $U_0$ in (a) being shown in
arbitrary units. 
}
\label{fig1}
\end{figure}

\begin{figure}
\caption{
Time courses of 
(a) $Z_{\ell}$ and (b) $Z_g$ 
for $\beta_0=0.1$, $\beta_1=0$, $J=0$ and $N=100$,
}
\label{fig2}
\end{figure}

\begin{figure}
\caption{
(a) The $\beta_0$ dependence
and (b) the $\beta_1$ dependence
of $\delta t_{o\ell}$ (squares)
and $\delta t_{og}$ (circles) for 
$\beta_1=0$ in (a) and $\beta_0=0.1$ in (b)
with $J=0$ and $N=100$,
filled symbols denoting results in DMA
and open symbols those in simulations.
}
\label{fig3}
\end{figure}

\begin{figure}
\caption{
Log-log plots of $\delta t_{o\ell}$ (squares)
and $\delta t_{og}$ (circles) against $N$ 
for (a) $\beta_1=0$ and (b) $\beta_1=0.05$ 
with $\beta_0=0.1$ and $J=0$,
filled symbols denoting results in DMA
and open symbols those in simulations.
}
\label{fig4}
\end{figure}

\begin{figure}
\caption{
The $J$ dependence of $\delta t_{o\ell}$ (squares)
and $\delta t_{og}$ (circles) 
for (a) $\beta_1=0$ and (b) $\beta_1=0.05$
with $\beta_0=0.1$ and $N=100$,
filled symbols denoting results in DMA
and open symbols those in simulations.
}
\label{fig5}
\end{figure}

\begin{figure}
\caption{
The time course of synchronization ratio $S$
for (a) $\beta_0=0.1$, $\beta_1=0$ and $J=100$,
(b) $\beta_0=0.1$, $\beta_1=0$ and $J=200$, and
(c) $\beta_0=0.1$, $\beta_1=0.05$ and $J=100$
with $N=100$,
solid curve denoting results of DMA and 
dashed curve those of simulations.
}
\label{fig6}
\end{figure}

\begin{figure}
\caption{
The dependence of the maximum of the synchronization ratio
on (a) $J$, (b) $N$, (c) $\beta_0$ and (d) $\beta_1$:
filled and open circles denote $S_{max}$ of DMA and 
simulations, respectively:
filled squares express $S'_{max}$ of DMA (see text). 
}
\label{fig7}
\end{figure}

\begin{figure}
\caption{
Time courses of (a) $\mu_v$ and (b) $\gamma_{v,v}$ 
for Poisson spike inputs
with the average ISI of 25 ms for $\beta_0=0.1$,
$\beta_1=0$, $J=0$ and $N=0$,
solid and dashed curves in (a) denoting results
of DMA and simulations, respectively.
$K^{(e)}$ and $U_o$ in (a) is plotted in arbitrary units.
The result of simulations in (b) is shifted upwards
by 30.
}
\label{fig8}
\end{figure}

\begin{figure}
\caption{
Time courses of (a) $\mu_v$ and 
(b) $\sigma_{\ell}\:(=\surd{\gamma_{v,v}})$
with $\beta_0=0.1$, $\beta_1=0$ and $N=1$
for constant current input of $I_{i}=20$,
solid, dotted and dashed curves denoting
results of DMA, 
DMA2 (the second-order DMA) and
simulations, respectively.
A constant input current is shown at the bottom of (a)
(see text).
}
\label{fig9}
\end{figure}

\begin{figure}
\caption{
Time courses of (a) $\mu_v$ and 
(b) $\sigma_{\ell}\:(=\surd{\gamma_{v,v}})$
with $\beta_0=0.2$, $\beta_1=0$ and $N=1$
for a periodic spike train input with ISI of 25 ms, 
solid, dotted and dashed curves denoting
results of DMA, 
DMA2 (the second-order DMA) and
simulations, respectively.
A periodic input spike is shown at the bottom
of (a) (see text).
}
\label{fig10}
\end{figure}

\end{document}